\documentclass[preprint]{aastex6}

\usepackage{slashbox}

\newcommand\erf{\rm erf}

\newcommand{\be}{\begin{equation}}
 \newcommand{\ee}{\end{equation}}
 \newcommand{\au}{{\rm au}}
 
\def\brk#1{{\left[#1\right]}}
\def\prn#1{{\left(#1\right)}}

\begin{document}

\title{Exploring the Grain Properties in the Disk of HL Tau with an Evolutionary Model}
\author{Carlos Tapia \altaffilmark{1},  Susana Lizano \altaffilmark{1}, Anibal Sierra \altaffilmark{1}, Carlos Carrasco-Gonz\'alez\altaffilmark{1}, 
and Elly Bayona-Bobadilla \altaffilmark{1,2} }

\altaffiltext{1}{Instituto de Radioastronom\'ia y Astrof\'isica, Universidad Nacional Aut\'onoma de M\'exico, Apartado Postal 3-72, C.P. 58089 Morelia, Michoac\'an, M\'exico}\
\altaffiltext{2}{Instituto de Ciencias Nucleares, Universidad Nacional Aut\'onoma de M\'exico, Apartado Postal 70-468, C.P. 04510 Cd. Mx., M\'exico}
\

\begin{abstract}
We model the ALMA and VLA millimeter  radial profiles of the disk around HL Tau to constrain the properties of the dust grains.
We adopt the disk evolutionary models of Lynden-Bell \& Pringle and calculate their temperature and density structure and emission. These disks are heated by the internal viscosity and irradiated by the central star and a warm envelope.  We consider a dust size distribution $n(a) da \propto a^{-3.5} da $, and vary the maximum grain size in the atmosphere and the midplane, $a_{\rm max}=100\ \mu$m, 1 mm, and 1cm. We also include dust settling and vary the dust-to-gas mass ratio from 1 to 9 times the ISM value. We find that the models that can fit the observed level of emission along the profiles at all wavelengths have an atmosphere with a maximum grain size  
$a_{\rm max} = 100 \ \mu$m, and a midplane with $a_{\rm max}=1$ cm.  The disk substructure, with a deficit of emission in the gaps, can be due to dust properties in these regions that are different from those in the rings. We test an opacity effect (different $a_{\rm max}$) and a dust mass deficit (smaller dust-to-gas mass ratio) in the gaps. 
  We find that  the emission profiles are better reproduced by models with a dust deficit in the gaps, 
 although a combined effect is also possible. These models have a global dust-to-gas mass ratio  twice the ISM value, needed to reach the level of emission of the 7.8 mm VLA profile. 
\end{abstract}


\section{Introduction}
\label{sec:intro}
The HL Tau disk was one of the first sources where a high angular resolution map revealed a disk with multiple rings and gaps \citep{ALMA_2015}. Many physical models have been proposed to explain the structure observed in this disk, these include planets in the gaps \citep{Jin_2016}, three planets and disk evolution \citep{DiPierro_2015}, a super-earth and a low viscosity disk \citep{Dong_2018}, dust radial migration and ice lines \citep{Okuzumi_2019}, and gaps at the edge of the disk dead zone 
\citep{Flock_2015}. In addition, several other papers have inferred dust physical properties 
(maximum grain size, dust temperature, and optical depth) by modelling the observed multi-wavelength 
emission either by fitting the disk parameters or by calculating the structure of passive disks irradiated by the central star
(e.g. \citealt{Kwon_2015}; \citealt{Carrasco-Gonzalez_2016}; \citealt{Pinte_2016}; \citealt{Liu_2017};
 \citealt{Carrasco-Gonzalez_2019}, hereafter CG19). All these latter investigations infer millimeter dust grains in the HL Tau 
 disk based on the observed small spectral indices.
These millimeter values differ from the maximum grain size of some hundreds of micrometers required to explain the observed dust polarized emission (e.g., {\citealt{Kataoka_2015}; \citealt{Kataoka_2017}}).  It is not clear yet what is causing the disagreement between the estimates of the dust grain sizes. One possibility is that the polarized emission observed at ALMA wavelengths is 
mainly tracing the dust grains in the disk atmosphere, while the longer VLA  wavelengths trace deeper regions of the disk, closer to the midplane, where larger grains are expected to lie due to dust settling.  
Another possibility is a different dust composition as discussed
recently by  \cite{Yang_2019} who found that
 dust grains of absorptive carbonaceous material and a distribution with a maximum grain size of 3 mm can produce both the
 observed level of polarization and small spectral indices. It is therefore important to study the emission of disks with different dust properties and compare with multi-wavelength observations, to improve our understanding of their physical properties.

Knowledge of the dust physical properties and their evolution  is fundamental to explain the structure of the disks observed at millimeter wavelengths. Recently, ALMA high angular resolution observations of 20 nearby protoplanetary disks in the DSHARP project exposed the prevalence of different  morphologies  and substructure such as vortices, spiral arms, and rings \citep{Andrews_2018}. Hopefully, in a near future one can obtain
high angular resolution multi-wavelength information of all these sources to be able to obtain detailed diagnostics of the dust properties in 
these disks.
 
In this work, we study the HL Tau disk using the viscously evolving disk  models of  \cite{Lynden-Bell_1974} with the age of the HL Tau system. We calculate the emission of these disks with different dust properties and compare it with the observed millimeter emission. 
To do this we calculate the radial and vertical structure of these models heated by the internal viscosity, and irradiated by both the central 
source and a warm  envelope (\citealt{Lizano_2016};  \citealt{Tapia_2017}). We then compare the azimuthally averaged intensity profiles 
of the models with 4 observed ALMA and VLA millimeter profiles in order to constrain the physical properties of the dust in the disk. 

This paper is organized as follows: Section \ref{SEC:Evolution} briefly describes the Lynden-Bell \& Pringle evolutionary
disk models. Section \ref{SEC:settling} explains how the dust settling is included in these models. Section \ref{SEC:Parameters} summarizes the star and disk parameters of the fiducial model, while Section \ref{SEC:Observations} describes the observed 
 millimeter profiles.
In Section \ref{SEC:Level} we obtain the emission of the models at millimeter wavelengths, to find the dust properties that can produce the observed level of emission along the profiles. Next, in  Section \ref{SEC:Structure}, 
we explore the effect of changing the dust properties in the gaps in order to reproduce the observed bright-dark ring substructure 
of the HL Tau disk. In Section \ref{SEC:Settling2} we discuss the effect of changing the parameters of the fiducial model: the dust settling parameters, the initial disk mass, and the initial accretion rate. The discussion and conclusions are presented in Sections \ref{SEC:Discussion} and \ref{SEC:Conclusions}, respectively.

\section{Evolutionary disk model}
\label{SEC:Evolution}

The radial structure of viscously evolving accretion disks  with a power-law viscosity, $\nu \propto R^{\gamma}$,
was found by \cite{Lynden-Bell_1974}.  Due to the viscous evolution, 
the  disk surface density and mass accretion rate decrease with time. 
Here we will study models with a viscosity exponent $\gamma=1$
which can fit  the decline of the mass accretion rate with time for T Tauri stars with different ages 
(\citealt{Hartmann_1998}; Manzo-Mart\'\i nez et al. in prep.). Following the normalization of  \cite{Hartmann_1998}
the viscosity can be written as
\begin{equation}
    \nu = \nu_1 \left(\frac{R}{R_1}\right),
    \label{eq:nu}
\end{equation}
where $R_1$ is a normalization radius, and $\nu_1$ is the viscosity at $R_1$.  
The gas surface density is a power law with an exponential cut-off given by 
\begin{equation}
    \Sigma(R,{\mathbb T})=  \frac{M_{d0} }{2 \pi R_1^2} \left(\frac{R_1}{R}\right) {\mathbb T}^{-3/2}\exp \bigg(\frac{-R/R_1}{ \mathbb T}\bigg),
    \label{eq:sigma} 
\end{equation}
and the disk mass is given by 
\begin{equation}
    M_d(R,{\mathbb T})= \frac{M_{d0}}{{\mathbb T}^{1/2} }\Bigg[ 1 - \exp \bigg(\frac{-R/R_1}{ \mathbb T}\bigg) \Bigg] ,
    \label{eq:mass}
 \end{equation}
   where $M_{d0}$ is called the initial mass of the disk, 
\footnote{ Note that, for a finite disk with radius $R_d$, the initial disk mass is $ M(R_d, 0) = M_{d0}  \left\{ 1 - \exp{\left(-R_d/R_1\right)}\right\}$. }
and  the normalized time is
\begin{equation}\label{T}
   {\mathbb T}= \frac{t}{t_\nu} +1, 
   \label{eq:TT}
\end{equation}
where the viscous time is $t_\nu = R_1^2/(3\nu_1)$.
%From eq. (\ref{eq:mass}), for an initial disk radius $R_d$, $M(R_d, 0)$ is the initial disk mass and $M_d(0) = M(R_d, 0)/ \left\{ 1 - \exp{\left(-R_d/R_1\right)}\right\}$.}
%\footnote{The normalization radius $R_1$ measures the initial mass concentration of the disk, around 63 \% of $M_{d0}$ is within this radius.}
In addition, the mass accretion rate  is given by
\begin{equation}
    \dot M_d(R,\mathbb T) = \frac{\dot M_{*0} }{{\mathbb T}^{3/2}}\ \exp \bigg(\frac{-R/R_1}{ \mathbb T}\bigg) \left[ 1 - \frac{2 R/ R_1}{\mathbb T} \right],
    \label{eq:dotM}
\end{equation}
where the initial mass accretion rate at the center is 
\begin{equation}
\dot M_{*0} \equiv \frac{M_{d0}}{2 t_\nu}.
\label{eq:mdots}
\end{equation}
For a disk with a given age $t=t_{\rm age}$, the normalized time (Equation \ref{eq:TT}) can be written as
\begin{equation}
   {\mathbb T}= \frac{t_{\rm age} }{t_\nu} +1 = 2 t_{\rm age} \left(\frac{ \dot M_{*0}}{M_{d0}}\right) +1 .
%  = 2 \times 10^6  \, \left(\frac{t_{\rm age} }{\rm Myr} \right) 
%  \left( \frac{ \dot M_*(0) }{M_\odot {\rm yr}^{-1}} \right) \left(\frac{M_d(0)}{M_\odot}\right)^{-1}+1,
\label{eq:T}
\end{equation}
In addition, one can define a characteristic radius 
\begin{equation}
    R_c = R_1 {\mathbb T},
\label{eq:Rc}
\end{equation}
where  the exponential term in the surface density starts to dominate (see Equation \ref{eq:sigma}). 

In terms of the normalized radius $r= R/ R_c$,
the gas surface density can be written as
\begin{equation}
    \Sigma(r,{\mathbb T})=  \frac{M_{d0} }{2 \pi R_c^2}  {\mathbb T}^{-1/2}   r^{-1} \exp \left( - r \right),
    \label{eq:sigma_mod} 
\end{equation}
and, and using Equation (\ref{eq:mdots}),  the viscosity in Equation (\ref{eq:nu}) can be written as
\begin{equation}
       \nu(r) = \frac{R_1^2}{3 t_\nu} \left(\frac{R}{R_1}\right)=\frac{2}{3} \frac{  \dot M_{*0} R_c^2}{ M_{d0} }\, 
   {\mathbb T}^{-1} r  .
    \label{eq:nu_dim}
\end{equation}
Given the age of the source $t_{\rm age}$, the surface density and viscosity 
depend only on  the initial disk mass $M_{d0}$, the initial mass accretion rate at the center $\dot M_{*0}$, 
and the characteristic radius $R_c$. These two functions are needed to determine the disk vertical structure and emission. 

\section{Dust-to-gas mass ratio and dust settling}
\label{SEC:settling}

The dust-to-gas mass ratio $\zeta$ will be described in terms of the dust ratio $\epsilon= \zeta/ \zeta_{\rm ISM}$, where
$\zeta_{\rm ISM} = 1/100$ is the ISM value.
The dust and gas masses in protoplanetary disks can be obtained from millimeter continuum emission and 
CO isotopologue lines, respectively.
Recently, \cite{Ansdell_2016} made a survey of the dust and gas content of young protoplanetary disks in the Lupus  
star forming region with ALMA. The lower panel of their Figure 3 shows that gas-to-dust mass ratio for all the sample. %, where  the ISM  value is $1/\zeta_{\rm ISM}$ = 100. 
The observed ratios correspond to values of the dust ratio
  $\epsilon \sim 0.1 - 10$.
Also, \cite{Wu_2018} studied the gas and dust content of the disk and envelope of HL Tau with lower 
resolution SMA observations. Since they do not resolve the disk, they modeled the continuum,  
$^{13}CO$, and $C^{18}O$ observations. They estimate very low values of the disk gas-to-dust mass ratio, in the range 0.07 to 4, 
which correspond to $\epsilon \sim 25 - 1400$ (see their Table 5, where they also summarize the results for other
sources). It would be important to estimate the gas and dust content of HL Tau with high angular resolution studies
to confirm these results. 
 In this work,  we will consider $\epsilon \ge 1$ as a model parameter to fit the observed ALMA and VLA dust continuum emission.

 We also include the process of dust settling and grain growth in the disk models. Following \cite{DAlessio_2006},
 the disk has two regions: the atmosphere with small grains, and the midplane   with bigger grains. 
 We assume a dust grain size distribution given by $n(a) d a \propto a^{-3.5} da $ \citep{Mathis_1977} with a minimum grain size 
$a_{\rm min} = 0.005 \ \mu$m, and  vary the maximum grain size $a_{\rm max}$, with 
 different values of $a_{\rm max}$ for the grain population in the atmosphere and in the midplane.
 % One can define the gas surface density that corresponds to the midplane and atmosphere as 
% $\Sigma_{\rm big} = \int_0^{z_{\rm big}} \rho dz$ and  $\Sigma_{\rm small} = \int_{z_{\rm big}}^\infty \rho dz$, respectively, where
% $\rho$ is the gas density, and $z_{\rm big}$ is the is the height of the midplane. 
% Then, total mass surface density is  $\Sigma(R) = 2 (\Sigma_{\rm small} + \Sigma_{\rm big}) $.
   At each radius, the disk has a total dust mass determined by the  dust-to-gas mass ratio $\zeta(R)$. This radial function can be used to 
   take into account, for example, dust migration, growth, and sintering (e.g., \citealt{Brauer_2008}; \citealt{Okuzumi_2016}). One assumes for the atmosphere
   a dust-to-gas mass ratio  $\zeta_{\rm small}$ that takes into account the dust mass lost by settling to the midplane. 
   The midplane has  a  dust-to-gas mass ratio  $\zeta_{\rm big}$ that takes into account the dust mass gained from the atmosphere. 
  The surface density of the midplane is defined as  $\Sigma_{\rm big} = 2 \int_0^{z_{\rm big}} \rho dz$ where $z_{\rm big}$ 
   is the height of the midplane region and $\rho$ is the gas volume density.
    %  The total dust mass at each
 %  radius is conserved, so that $\zeta(\varpi) \Sigma(R) = \zeta_{\rm small} \Sigma_{\rm small} + \zeta_{\rm big} \Sigma_{\rm big}$.
    The degree of dust settling is determined by the dust ratio  $\epsilon_{\rm small} = \zeta_{\rm small}/ \zeta_{\rm ISM}$.
    %, where $\zeta_{\rm ISM} = 1/100$ is the ISM value. 
    %From the condition of dust mass conservation one can derive the value of $\zeta_{\rm big}$. 
    Given  the parameters $\epsilon_{\rm small}$ and  $\Sigma_{\rm big}$, 
  Appendix \ref{App:1} shows the derivation of dust ratio  in the midplane $\epsilon_{\rm big} =  \zeta_{\rm big}/ \zeta_{\rm ISM}$.
  % in terms of the mass surface density measured from the midplane $\sigma =  2 \int_0^{z} \rho dz$.
 
\section{Parameters of the star-disk model}
\label{SEC:Parameters}

The values of the stellar and disk parameters of HL Tau are shown in Table \ref{Table:Parameters}:
the stellar mass, age, temperature, and radius, the initial disk mass and mass accretion rate, the disk radius, and the characteristic radius.   
For the initial disk mass we choose  $M_{ d 0} = 0.55 M_\odot$, less than the upper mass limit  
 for disk stability,  $M_d^{\rm max}  \sim M_*/3$ \citep{Shu_1990}.   The initial mass accretion rate 
 $\dot M_{*0} = 5 \times 10^{-7} M_\odot {\rm yr^{-1}}$ was obtained from Equations 
 \ref{eq:dotM} and \ref{eq:T}, given the actual mass accretion rate of HL Tau $\dot M_{*} = 10^{-7} M_\odot {\rm yr^{-1}}$ measured by  \citep{White_2004}.
  The central luminosity is the stellar plus accretion luminosities, $L_{\rm c} = L_* + \eta (G M_* \dot M_*)/ R_*$,
  where the factor $\eta=(1-R_*/R_{\rm int})$ takes into 
  account the disk truncation by the stellar magnetosphere 
and  $R_{\rm int} \sim  5 R_*$ \citep{Gullbring_1998}.  
For the parameters in Table \ref{Table:Parameters}, 
 $L_{\rm c} = 10.7 \ L_\odot$,  which is within the luminosity range inferred by  \cite{Robitaille_2007} 
from the SED modelling of HL Tau. We will assume a distance to the HL Tau of 147 pc \citep{Galli_2018}. 

  \vfill\eject
  
\begin{deluxetable}{cccccccc}
\tablecolumns{8}
\tablewidth{0pc}
%\tabletypesize{\scriptsize}
\tablecaption{\label{Table:Parameters} HL Tau Stellar and Disk Parameters}
\tablehead{
 \colhead{$M_{*} \, ^{\rm a} $}   & \colhead{age$\, ^{\rm b} $}& \colhead{${T_*}\, ^{\rm c}$ } & \colhead{$ {R_*}\, ^{\rm c}$} 
 & \colhead{$M_{d0}\, ^{\rm d} $} & \colhead{$\dot M_{*0}\, ^{\rm e}$ } &  \colhead{$ R_d \, ^{\rm f}$}  & \colhead{$R_c \, ^{\rm g}$}
 \\
$(M_\odot)$   &  $({\rm Myr})$ &  (K)  & $(R_\odot)$  & $(M_\odot)$ 
 & $(M_\odot \, {\rm yr}^{-1})$   & $({\rm au})$  &  $({\rm au})$  
}
\startdata
1.7 &     1 &  4615 &  3.2 & 0.55 &   $ 5 \times 10^{-7} $  & 150  &  80
\enddata 
\tablecomments{ (a) \cite{Pinte_2016};  (b) \cite{VanderMarel_2019}; (c) from 
 \cite{Siess_2000} evolutionary tracks of low and intermediate mass stars;
 (d) maximum stable disk mass; (e) initial mass accretion rate that corresponds to the measured mass accretion rate of HL Tau by \cite{White_2004}; (f) \cite{Pinte_2016}; (g) \cite{Kwon_2011}.
 }
\end{deluxetable}

We first analyze a disk model with the parameters shown in this table.
This model has  the  maximum  gas surface density to be able to reproduce the level of the 
 7.8 mm VLA profile (see next section). The current mass of the disk  is  
 $M_d(R_d, 1 \ {\rm Myr}) = 0.28 M_\odot$  (see Equation \ref{eq:mass}).
 We further assume that the dust is settled in the midplane with a dust ratio in the disk atmosphere $\epsilon_{\rm small} = 0.1$ and 
a normalized gas surface density of the midplane $s_{\rm big} = \Sigma_{\rm big}/\Sigma(r,{\mathbb T}) =0.4$;
where $\Sigma(r, {\mathbb T})$ is given by Equation \ref{eq:sigma_mod}. For an isothermal disk, $s_{\rm big} = 0.4 $ corresponds to 
a settling height  
$z_{\rm big} = 0.53 \, H$, where $H$ is the gas scale height evaluated at the midplane temperature (see Appendix \ref{App:1}). 
 We call the model with the parameters in Table \ref{Table:Parameters} and  the dust settling parameters $\epsilon_{\rm small} =0.1$ 
and $s_{\rm big} = 0.4$, the  {\it fiducial model}. Below we will explore the effect of varying these parameters with respect to the fiducial model.

For the dust composition, we assume a dust mixture of silicates, organics, and ice with  a mass fractional abundances with respect to gas  $\zeta_{\rm sil} = 3.4 \times 10^{-3}$, 
$\zeta_{\rm org} = 4.1 \times 10^{-3}$, and $\zeta_{\rm ice} = 5.6 \times 10^{-3}$,
with bulk densities $\rho_{\rm sil} = 3.3 \, {\rm g \, cm}^{-3}$, $\rho_{\rm org} = 1.5 \, {\rm g \, cm}^{-3}$,
and $\rho_{\rm ice} = 0.92 \, {\rm g \, cm}^{-3}$
%, and  sublimation temperatures $T_{\rm sil, sub}=1400$K, $T_{\rm sil, sub}=425$K and $T_{\rm ice, sub}=180$K 
(\citealt{Pollack_1994}). We use the code of \cite{DAlessio_2001} which adopts the optical constants of  \cite{Draine_1984} for silicates, \cite{Warren_1984} for water ice, and \cite{Greenberg_1996} for organics.
We consider the total opacity due to both true absorption and scattering, $\chi_\nu = \kappa_\nu + \sigma_\nu$, 
where $\kappa_\nu$ and $\sigma_\nu$  are 
the absorption  and scattering mass mass coefficients, respectively.
For a given  maximum grain size $a_{\rm max}$ and  dust-to-gas mass ratio $\zeta = \epsilon  \zeta_{\rm ISM}$, 
one computes the mean and  monochromatic opacities 
to calculate the vertical structure and the emission of the models.
 
 In these models, the disk is heated by viscous dissipation, stellar and accretion irradiation, and
 a warm envelope \citep{DAlessio_1997}, and cools by dust emission. We assume the dust and the gas are 
 thermally coupled by collisions and have the same temperature.
 To obtain the vertical structure and emission of the disk, we follow the formalism of \cite{Lizano_2016}  
 without magnetic field, where the disk temperature $T$ and gas density $\rho$  %, and the radiation mean intensity $J$
 at each radius are calculated as a function of the mass surface density variable (see their Equations (11-13) and (17-18)),
 and the total disk flux has a viscous and a reprocessed component. The latter is due to the disk surface irradiation. 
 For the irradiation of the envelope
 we assume a thermal bath at the disk surface, with a temperature profile $T_e = T_0  \left( R /100  \,\au \right)^{-q}$  
(see \citealt{Tapia_2017}).
 %For the temperature of the envelope, we assume a radial profile $T_e = T_0  \left( R /100  \,\au \right)^{-q}$. 
 This warm envelope prevents the outer radii of the disks from becoming too cold and has been used to reproduce the
 millimeter SED and  profiles of HL Tau (e.g, \citealt{DAlessio_1997}; \citealt{Tapia_2017}).
 For all models we use $T_0 = 35 $ K  and  $q=0.16$ which is enough to heat the disk
outer regions. 

 \section{ The observed  millimeter profiles of HL Tau}
 \label{SEC:Observations}
 
We compare the emission of the models with HL Tau millimeter images at 
870 $\mu$m and 1.3 mm \citep{ALMA_2015}, and 2.1 mm (ALMA), and 
  7.8 mm (VLA) from  CG19.
The VLA image at $\lambda = 7.8$ mm is a combination between the Ka and Q bands, which are decontaminated from free-free emission.
The images at all wavelengths are convolved to an angular resolution of 50 m.a.s. (7.35 AU at 147 pc). 
The intensity profiles shown in Figures \ref{Fig:100mu}-\ref{Fig:MdMdot} below correspond to the azimuthally averaged 
intensity profiles of the disk maps taking into account the disk inclination. 
We find that with an inclination angle $i^{\rm model} = 37$ deg, the convolved model images shown below have
an aspect ratio between the major and minor axis that corresponds to the inclination inferred from the observed maps 
$i^{\rm obs} \sim 47$ deg.
Finally, because the signal-to-noise of the VLA 7.8 mm image is poor beyond 60 au, 
we will only compare the models emission within this radius.
%(i = 46.72 deg). 

 \section{The multi-wavelength emission level}
 \label{SEC:Level}
 
In this section, we aim to obtain the dust properties in the disk that can reproduce the level of emission along the 
observed profiles simultaneously at all 4 millimeter wavelengths. We will discuss the profiles substructure in the next section.

\begin{figure}
\centering
\includegraphics[scale=0.3]{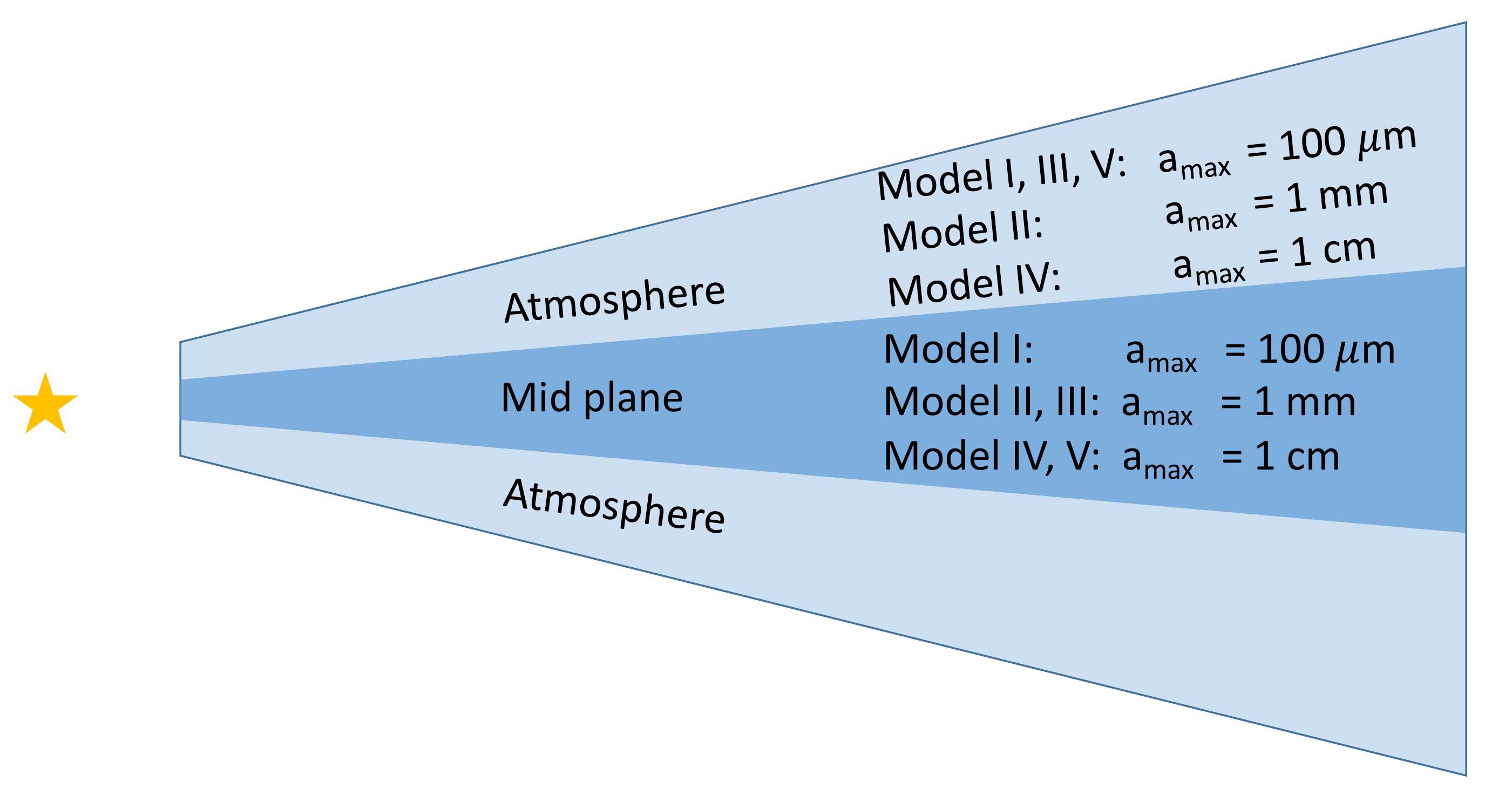}
\caption{ Sketch of the atmosphere and midplane of the different models in Table \ref{Table:Models}. }
\label{Fig:sketch}
\end{figure}

To compare the emission of the models with the observed azimuthally averaged ALMA and VLA profiles of HL Tau at 0.87 mm, 1.2 mm, 2.1mm, and 7.8 mm, we vary the maximum grain size a$_{\rm max}$ and the dust-to-gas mass ratio of the disk models. 
%In this section  we want to obtain the level of emission of the 4 wavelengths simultaneously. 
We explore models with
$a_{\rm max} = 100 \ \mu$m, 1 mm, and 1 cm. We also consider these different values of $a_{\rm max}$ 
in the disk atmosphere and the disk midplane.
In this section, we assume a dust-to-gas mass ratio $ \zeta= \epsilon \zeta_{\rm ISM}$ with a constant 
dust ratio  $1 \leq \epsilon \le 9$. As discussed above, such values for the dust ratio are found, 
for example, in protoplanetary disks in the Lupus star forming region \citep{Ansdell_2016}. 
Table \ref{Table:Models} shows the different models that we discuss below. Figure \ref{Fig:sketch} shows a sketch of the different models.
 
 \begin{deluxetable}{ccc}
 \tablewidth{10cm}
 \tablecolumns{3}
\tablecaption{\label{Table:Models} Models of Dust Size Distribution }
\tablehead{ 
      \colhead{Model }                       & \colhead{Atmosphere}   & \colhead{midplane} \\
               &  $a_{\rm max}$  &  $a_{\rm max}$ 
}
\startdata
I   &  100 $\mu$m & 100 $\mu$m \\
II    & 1 mm &  1 mm \\
III     & 100 $\mu$m  &  1 mm  \\
IV    & 1 cm  &  1 cm \\
V     &  100 $\mu$m   &  1 cm \\
\enddata
\end{deluxetable}

\begin{figure}
\centering
\includegraphics{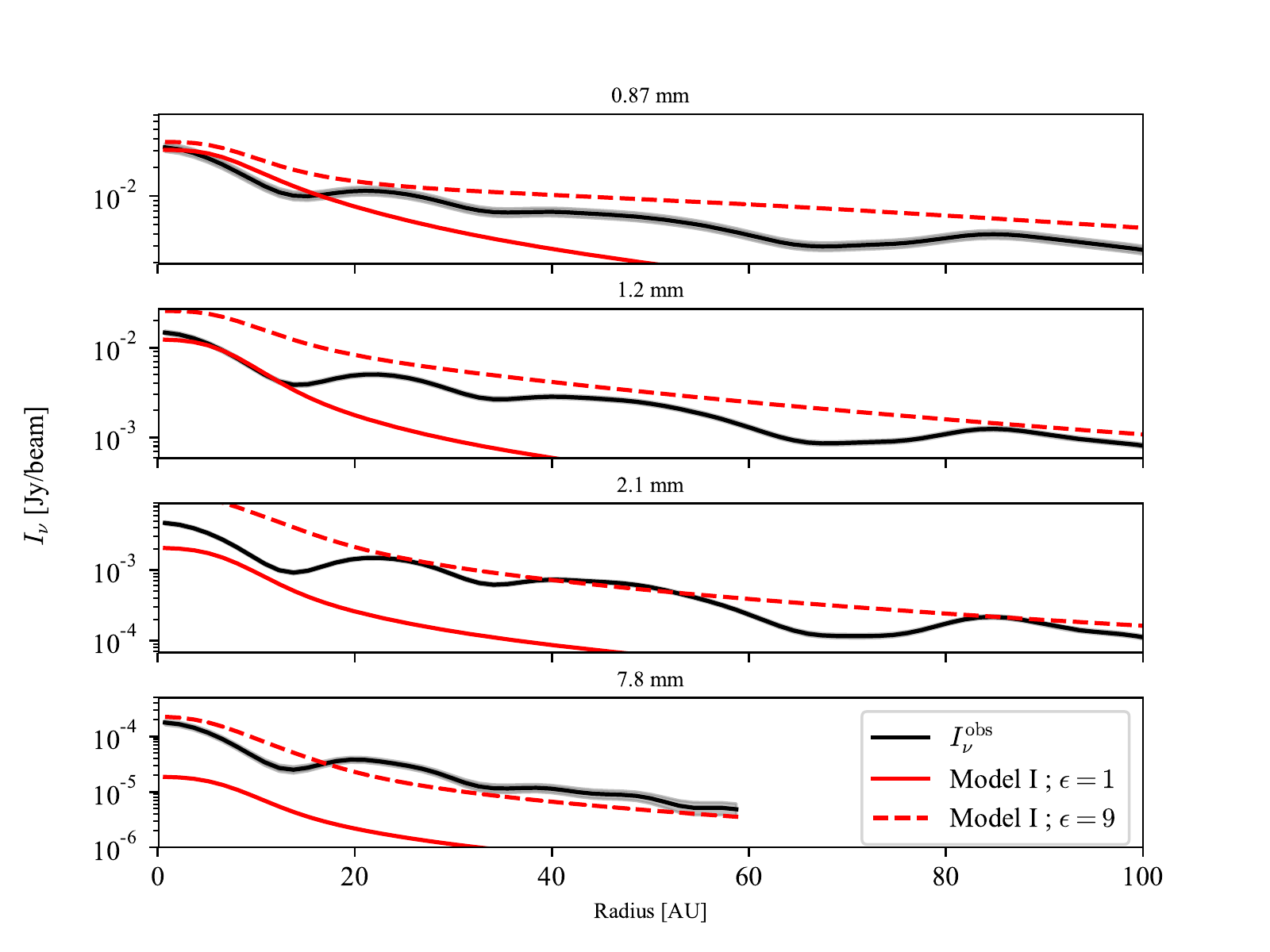}
\caption{The black solid lines show the  observed azimuthally averaged  profiles at 
each wavelength. The grey solid lines surrounding these profiles correspond to the error bars.
The red lines show the profiles of Model I with a dust-to-gas mass ratio $\zeta=\epsilon \zeta_{\rm ISM}$, with a dust ratio  $\epsilon=1$ (solid lines) and $\epsilon=9$ (dashed lines).}
\label{Fig:100mu}
\end{figure}

Figure \ref{Fig:100mu}  shows as black solid lines the observed azimuthally averaged  profiles at 
each wavelength indicated on the top of each panel: 0.87 mm, 1.2 mm, 2.1mm, and 7.8 mm, from top to bottom. The grey solid lines
surrounding these profiles correspond to the error bars that 
take into account  the  RMS and systematic errors due to the flux calibration
% $\Delta I_\nu = \sqrt{ \sigma_\nu^2 + (\eta_\nu I_\nu)^2 } $,
% where $ \sigma_\nu = \sigma^{\rm RMS}/ \sqrt{N}$, and $\eta_\nu = 0.1$ for 0.87 $\mu$m and 7.8 mm,  and $\eta=0.05$ for 1.2 mm and 2.1 mm 
(see detailed discussion in CG19).
The red lines show the profiles of
Model I: a disk with $a_{\rm max} = 100 \ \mu$m in both the atmosphere and the midplane. 
The solid red lines correspond to  a dust-to-gas mass ratio $\zeta=\epsilon \zeta_{\rm ISM}$ with a dust ratio  $\epsilon=1$; 
the red dashed lines correspond to a dust ratio   $\epsilon=9$. This high value of the 
 dust ratio can reproduce the level of emission at all 4 wavelength
 in the external part of the disk ($R > 40$ au) although the emission is a bit low at 7.8 mm. 
 Moreover, the model profiles at 1.2 mm and 2.1 mm 
 have too much emission at the inner region.

\begin{figure}
\centering
\includegraphics{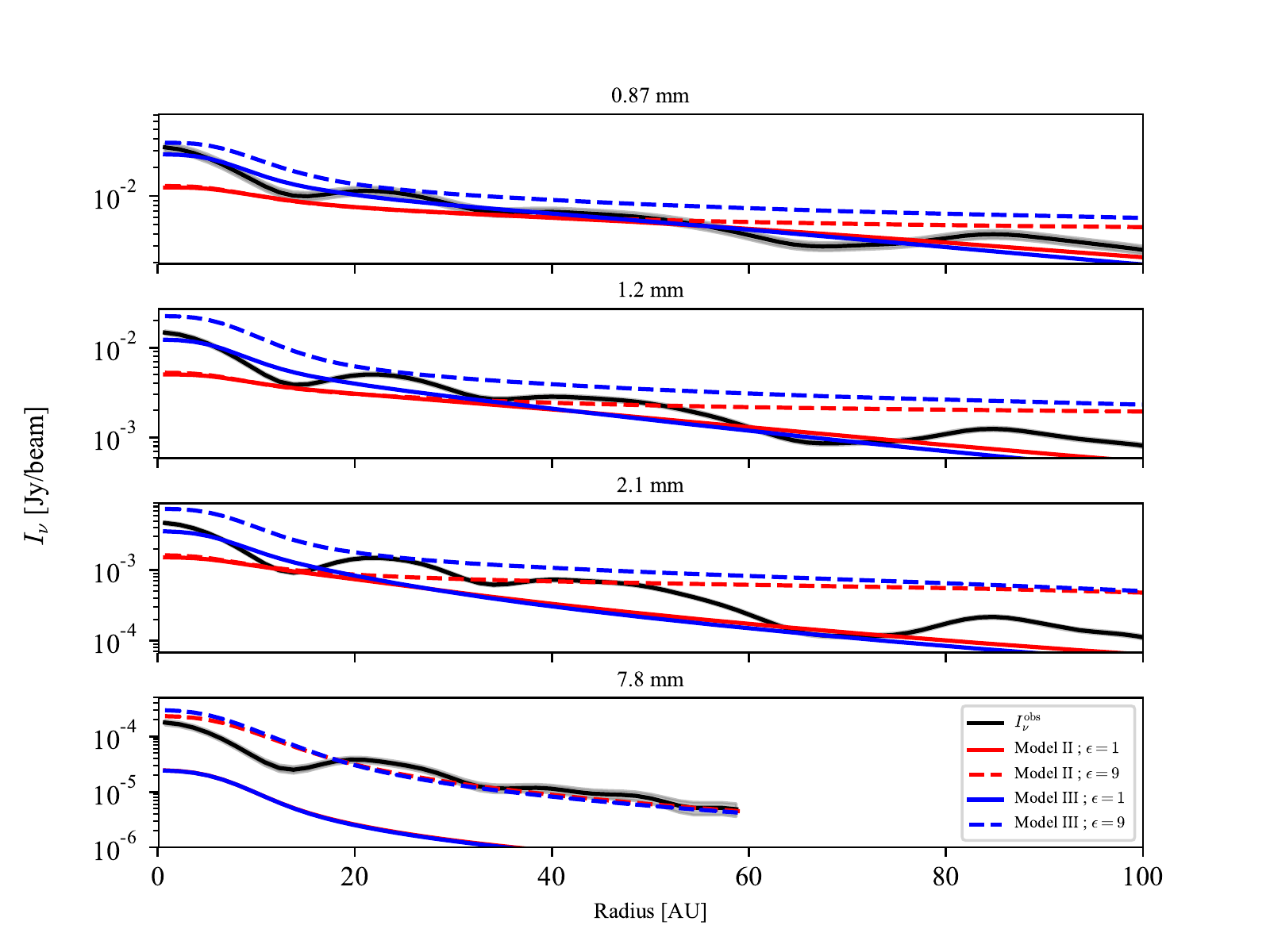}
\caption{
 The black solid lines show the observed  profiles at  each wavelength.
The red  lines show the profiles of
Model II with  a dust-to-gas mass ratio $\zeta=\epsilon \zeta_{\rm ISM}$,  
with a dust ratio  $\epsilon=1$ (solid lines) and  $\epsilon=9$ (dashed lines).
The blue lines show the profiles of
Model III with a dust ratio  $\epsilon=1$ (solid lines) and  $\epsilon=9$ (dashed lines).}
\label{Fig:1mm}
\end{figure}

Figure \ref{Fig:1mm} shows as black solid lines the observed profiles at each wavelength.
 Each panel shows 2 models, each with 2 values of the dust ratio  $\epsilon$.
The red  lines show the profiles of
Model II: a disk with $a_{\rm max} = 1$ mm in both the atmosphere and the midplane. 
The solid red lines correspond to  a
 dust-to-gas mass ratio $\zeta= \epsilon \zeta_{\rm ISM}$ with a dust ratio  $\epsilon=1$; the red dashed lines correspond to 
 a dust ratio  $\epsilon=9$.
The blue lines show the profiles of
Model III: a disk with $a_{\rm max} = 100 \ \mu$m in the atmosphere and 1 mm in the midplane.
The blue solid lines correspond to  a
 dust-to-gas mass ratio $\zeta=\epsilon \zeta_{\rm ISM}$ with a dust ratio  $\epsilon=1$; the blue dashed lines correspond to a 
 dust ratio   $\epsilon=9$. This large value of $\epsilon$ can reproduce 
 the level of emission at 7.8 mm but the models have too much emission for radii $R > 50$ au, especially the 2.1 mm profile.

\begin{figure}
\centering
\includegraphics{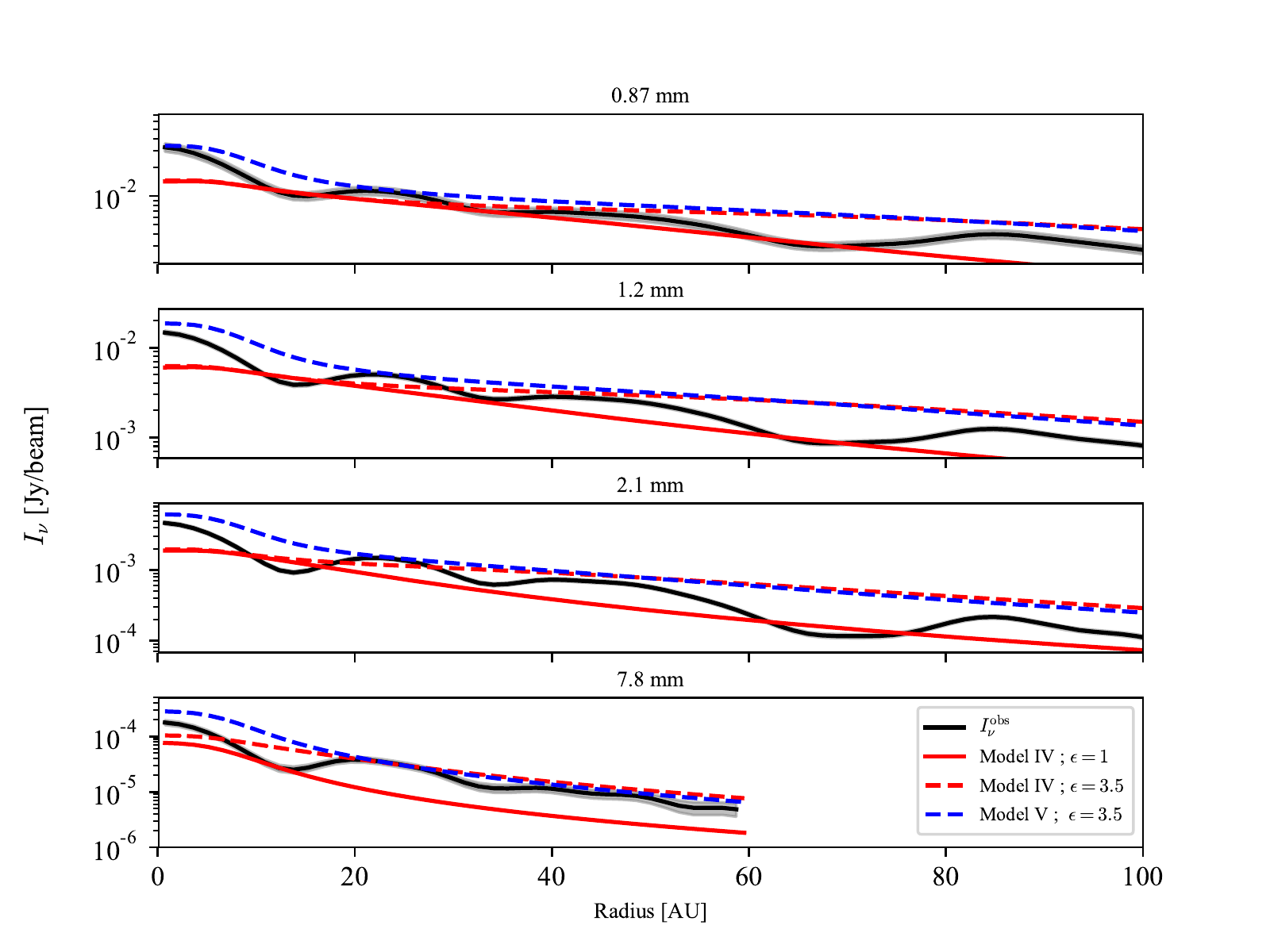}
\caption{The black solid lines show the observed profiles at at  each wavelength.
The red  lines show the profiles of
Model IV with a
 dust-to-gas mass ratio $\zeta=\epsilon \zeta_{\rm ISM}$, with a dust ratio  $\epsilon=1$ (solid lines), and
a  dust ratio  $\epsilon=3.5$ (dashed lines). The blue dashed lines 
 show the profiles of
Model V with a dust ratio  $\epsilon=3.5$.
 }
\label{Fig:1cm}
\end{figure}

Figure \ref{Fig:1cm} shows as black solid lines the observed profiles at each wavelength.
This figure also shows 2 models with different values of the dust ratio   $\epsilon$.
The red  lines show the profiles of
Model IV: a disk with $a_{\rm max} = 1$ cm in  both the atmosphere and the midplane. The solid red lines correspond to  a
 dust-to-gas mass ratio $\zeta=\epsilon \zeta_{\rm ISM}$ with a dust ratio  $\epsilon=1$; the red dashed lines correspond to a 
 a dust ratio  $\epsilon=3.5$. The blue dashed lines  show the profiles of
Model V, a disk with $a_{\rm max} = 100 \ \mu$m in the atmosphere and 1 cm in the midplane, with a dust ratio  $\epsilon=3.5$. 
\footnote{We omit Model V with  $\epsilon=1$ because it does not
 have enough emission at 7.8 mm.}
The latter model can reproduce the level of emission of the observed multi-wavelength profiles
with a moderate dust-to-gas mass ratio, a factor of 3.5 the ISM value. 
Thus, we will use this model to study the profiles  substructure in the next section.

\section{Disk substructure: rings and gaps}
\label{SEC:Structure}

The disk of HL Tau has a dark-bright ring substructure. In this section we explore if the fainter emission in
the gaps can be produced by changing the grain size or the amount of dust mass in 
the gaps with respect to the rings, in order to decrease the optical depth. 
\footnote{ We note that the temperature is not a free parameter of the model, 
but is obtained from local the processes of dust heating and cooling.}
In the first case, one considers a dust distribution in the gaps with  $a_{\rm max}$ smaller than in the rings.
In the second case,  one considers the same dust,  but a smaller dust ratio  $\epsilon$ in the gaps than in the rings.
To explore the effect of changing the dust properties in the gaps,  they are  located in the range of radii   6.5  $< R/{\rm au} < $ 16.5, 
31 $ < R/ {\rm au}  < $40, 60 $ < R/{\rm au} <$ 86, and $R/{\rm au} > 100$, 
consistent with the gap locations given by \cite{ALMA_2015}.

Differences in the dust grains sizes and/or dust mass between gaps and rings are expected, 
for example, when planets produce gaps where the dust is depleted and also induce pressure bumps at 
the gap edges. These pressure maxima act as dust traps where dust particles with large Stokes parameters accumulate,  while the smaller dust particles remain in the gaps, coupled with the gas (e.g., \citealt{Zhang_2018}).
Dust trapping is also produced in ring pressure maxima (e.g., \citealt{Sierra_2019}); or
outside the snow lines of different volatiles where sintering suppresses dust growth and produces a pile up of small dust  grains (e.g., \citealt{Okuzumi_2016}). Figure 7 of CG19 shows that the grain sizes and dust surface density in the gaps of HL Tau are smaller than in the rings. In their work the dust properties are inferred by modelling the 
radial dust spectral indices, assuming an isothermal vertical structure. Their inferred contrasts in dust properties 
between gaps and rings are limited by the spatial resolution of the observations (see the discussion in their Section 4.2).

\begin{figure}
\centering
\includegraphics{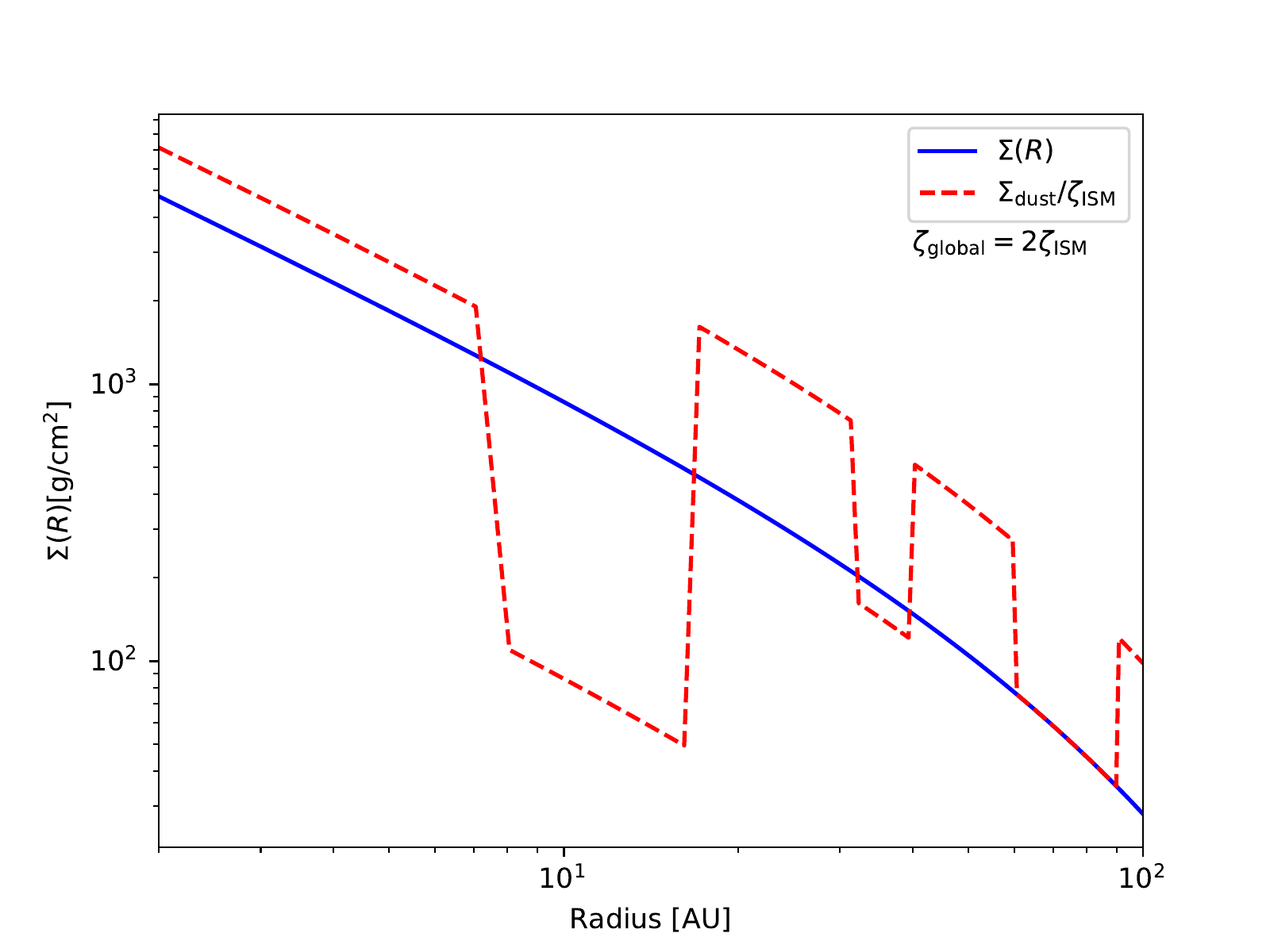}
\caption{The solid blue line shows the gas surface density radial profile $\Sigma(R)$ of the viscous disk. The dashed red line shows the 
normalized dust surface density radial profile of Model VII:  $\Sigma_{\rm dust}/ \zeta_{\rm ISM}$. 
The global dust-to-gas mass ratio of the disk
in this model is $\zeta_{\rm global} = 2 \ \zeta_{\rm ISM}$.
 }
\label{Fig:Sigma_dust}
\end{figure}

 Table \ref{Table:ModelVI} shows the parameters of Model VI which decreases the opacity in the gaps by decreasing the 
 size of the dust particles.
 This model has grains with $a_{\rm max} = 100 \, \mu$m  in the atmosphere;
 in the midplane it has grains with  $a_{\rm max} = 1 $ cm in the rings, and 
 grains with $a_{\rm max} = 100 \, \mu$m in the gaps.
 Table \ref{Table:ModelVII} shows the parameters of Model VII which has a deficit of dust mass in the gaps. 
 This model has  $a_{\rm max} = 100 \, \mu$m in the atmosphere and $a_{\rm max} =1$ cm in the midplane. 
 The dust ratio  is a function of radius, $\epsilon(R)$.
 Figure \ref{Fig:Sigma_dust} shows the dust surface density profile of this model.
 
\begin{figure}
\centering
\includegraphics{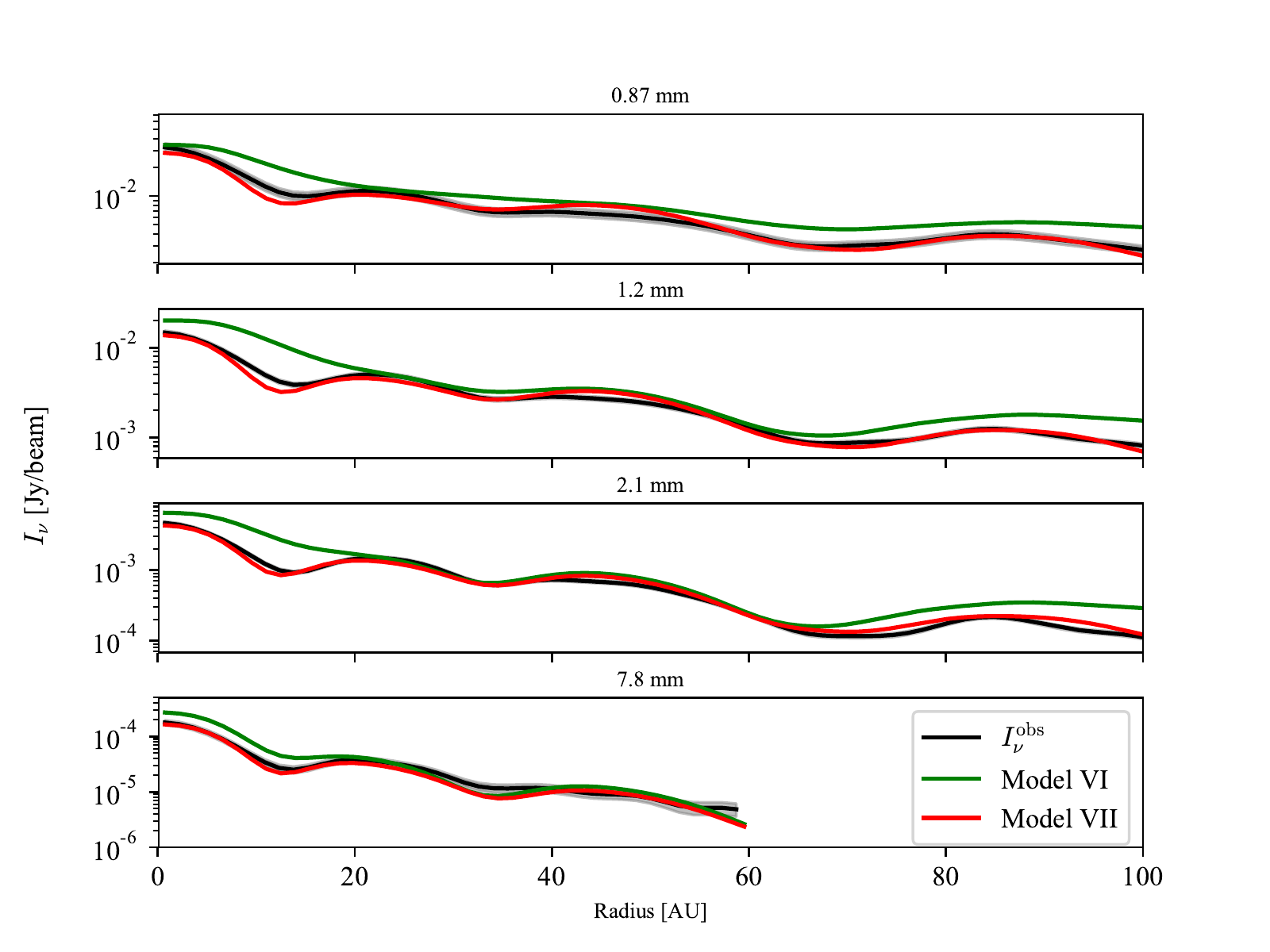}
\caption{The black lines show the observed profiles at at  each wavelength.
The green lines show the profiles of
Model VI.  The red lines show the profiles of Model VII.
 }
\label{Fig:1cm_gaps}
\end{figure}

  Figure \ref{Fig:1cm_gaps} shows the observed millimeter 
 profiles together with the profiles of models VI and VII. The green lines correspond to Model VI.
 The red lines correspond to Model VII. Model VI, which explores the opacity effect of small grains in the gaps, 
 has excess emission at the ALMA wavelengths in the first gap and beyond $\sim 60$ au.
  Model VII, which explores a mass deficit in the gaps, reproduces reasonable well the gaps and
 rings for the assumed function $\epsilon(R)$. A combination of both, a mass deficit and an opacity effect would also 
 reproduce the observed substructure.

  \begin{table}[t!]
\centering
\caption{Model VI  ($\epsilon=3.5$)}
\begin{tabular}{lc}
\hline
Atmosphere   &  $a_{\rm max} = 100 \ \mu$m  \\
Gaps midplane   &  $a_{\rm max} = 100 \ \mu$m  \\
Rings midplane     &  $a_{\rm max} = 1$ cm \\
\hline
\end{tabular}
\label{Table:ModelVI}
\end{table}

\begin{table}[t!]
\centering
\caption{Model VII ( $\epsilon(R))$}
\begin{tabular}{ccccccccc}
\hline 
   \multicolumn{9}{c}{Atmosphere:  $a_{\rm max} = 100\mu$m $\vert$  Midplane: $a_{\rm max} = 1 $ cm}     \\
   %     Model                         & Atmosphere   & midplane  &      \\
 %              &  $a_{\rm max}$  &  $a_{\rm max}$ \\
\hline
 $R$/au$^{\rm a} $ &  $<$ 7.5 & [ 7.5 - 16.5] & 16.5 - 32   & [32-40]  & 40 - 60   &[ 60 - 90]  & 90 - 100 & [$>$ 100 ] \\
$\epsilon(R)$   & 1.5 & 0.1 & 3.5 &  0.8 &  3.5& 1.0  & 3.5 & 1.5 \\
\hline
\end{tabular}
\tablecomments{ (a) Gaps are indicated with brackets.}
\label{Table:ModelVII}
\end{table}

\section{Changing the Disk Parameters}
\label{SEC:Settling2}

To explore the effect of changing the dust settling, we consider model VII with different settling parameters
$s_{\rm big} = 0.2, 0.4, 0.6$ and different dust ratios  $\epsilon_{\rm small}=0.1, 0.01$.
The settling parameters $s_{\rm big} = 0.2, 0.4, 0.6$ correspond to a dust scale height $\delta H$, with $\delta = 0.25, 0.53, $ and $ 0.84$, respectively (Appendix \ref{App:1}), where the gas scale height $H$ is evaluated at the midplane
temperature.

\begin{figure}
\centering
\includegraphics{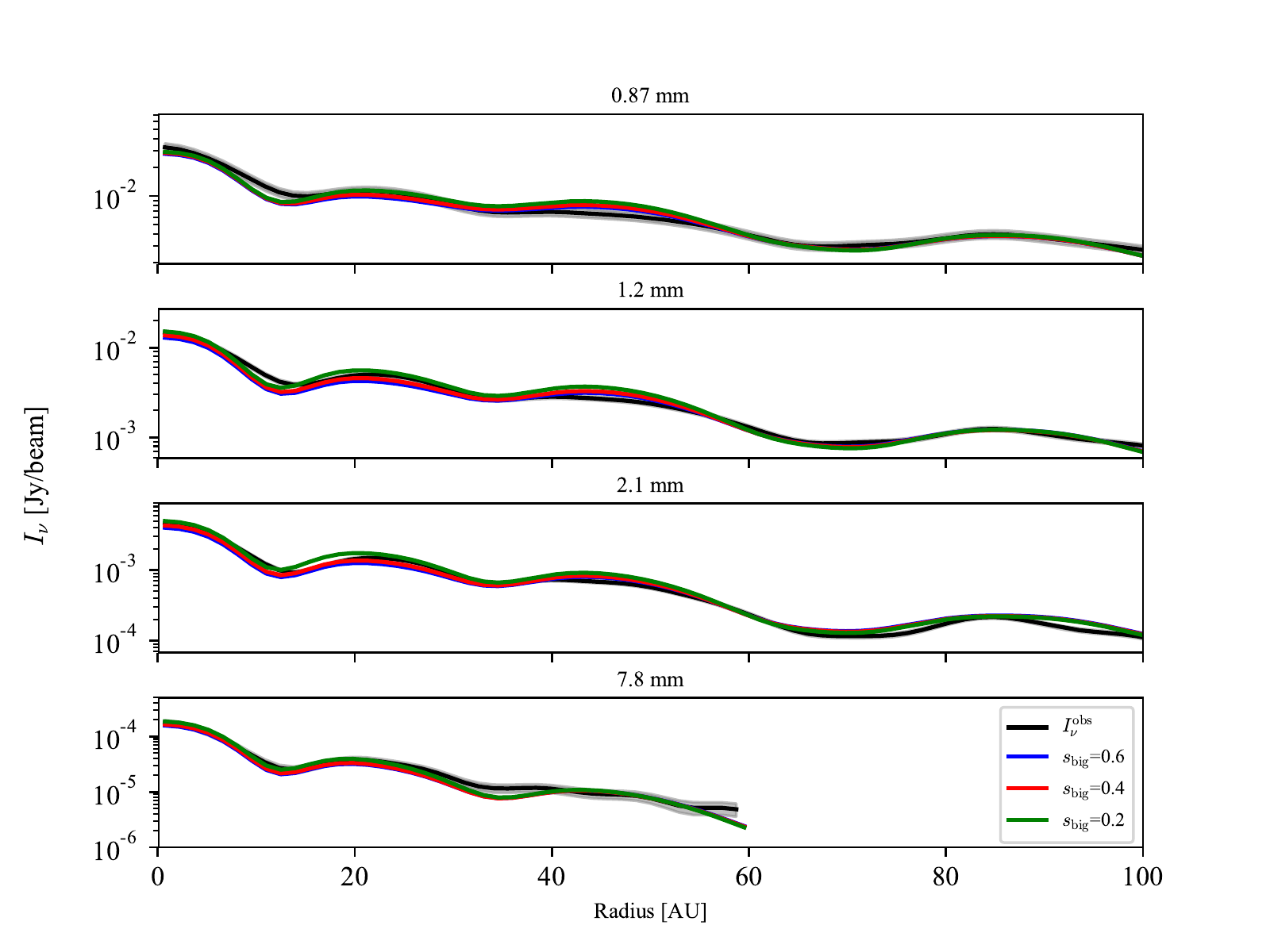}
\caption{The black solid lines show the observed profiles at at  each wavelength.
The red, green, and blue solid lines show the profiles of Model VI where $s_{\rm big} = 0.6, 0.4, 0.2$,
respectively. The red lines correspond to Model VII.  }
\label{Fig:Settling1}
\end{figure}

\begin{figure}
\centering
\includegraphics{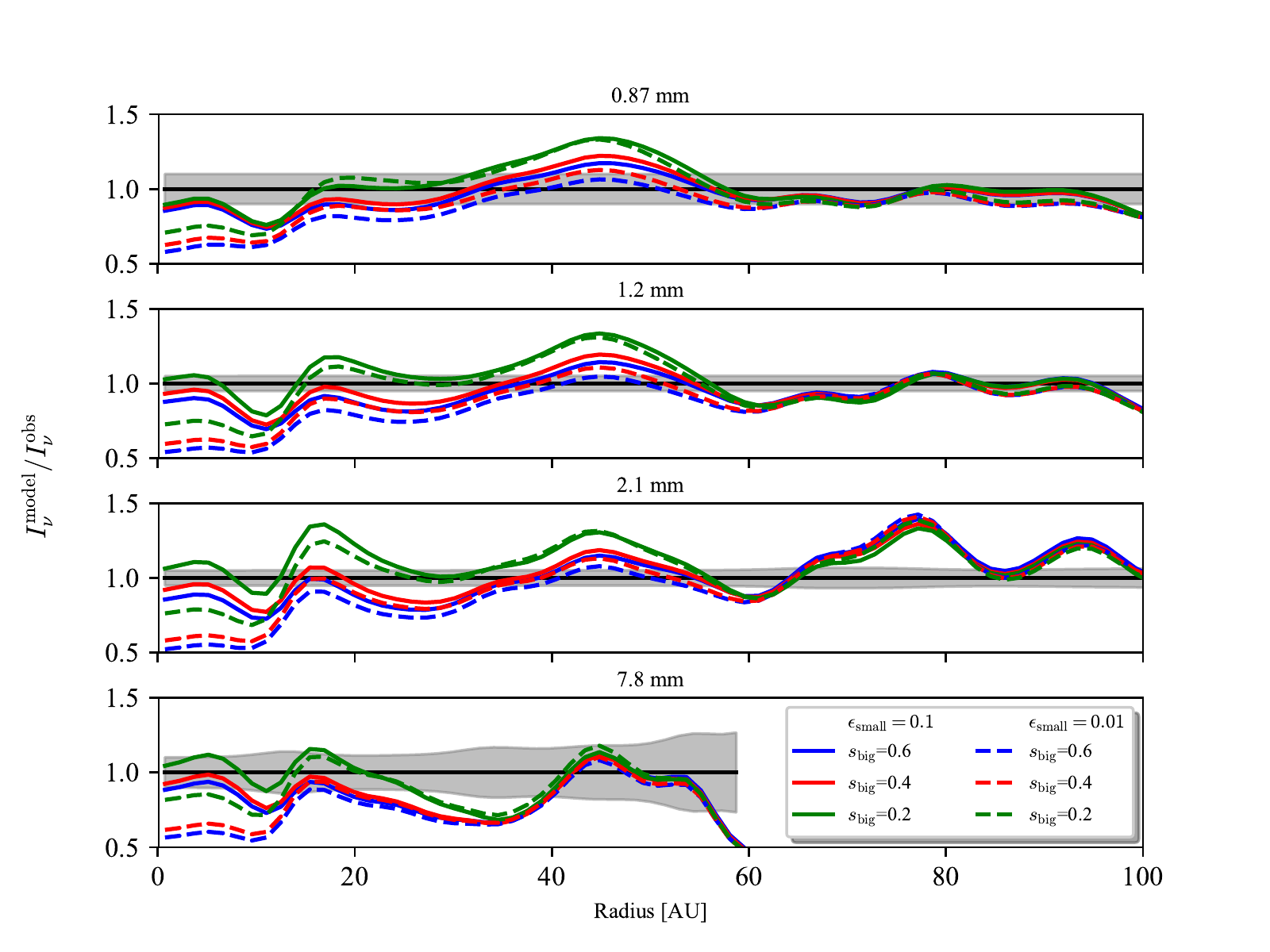}
\caption{The panels show the ratios of the model and observed intensities at each wavelength.
The grey shadows correspond to the error bars.
The red, green, and blue lines correspond to $s_{\rm big} = 0.6, 0.4,$ and 0.2, respectively.
The solid and dashed lines correspond to models with $\epsilon_{\rm small}=0.1$ and 0.01, respectively.
The solid red lines ($s_{\rm big}=0.4; \epsilon_{\rm small}=0.1$) correspond to Model VII.
 }
\label{Fig:Settling2}
\end{figure}

Figure \ref{Fig:Settling1}  shows the effect of varying $s_{\rm big}$ in the model profiles,
keeping the same $\epsilon_{\rm small}=0.1$. The red, green, and blue lines correspond to the model with
$s_{\rm big} = 0.6, 0.4, 0.2$, respectively, where $s_{\rm big} = 0.4$ corresponds to Model VII. 
Figure \ref{Fig:Settling2} shows the ratios of the model and the observed intensity at each wavelength. The code color
is the same as Figure \ref{Fig:Settling1}. The solid and dashed lines correspond to models with 
$\epsilon_{\rm small}=0.1$ and 0.01, respectively. 
The $\chi^2$ value of each model is calculated as
\begin{equation}
\chi^2  =  \sum_{\nu} \chi^2_\nu =  \sum_{\nu} \left[ \frac{1}{N_\nu}\sum_{j=1} ^{N_\nu} \frac{\left( I_\nu^{\rm model} - I_\nu^{\rm obs} \right)^2}{\left(I_\nu^{\rm obs}\right)^2} \right],
\end{equation}
where one sums over  $N_\nu$ radii. 
These values are shown in Table {\ref{Table:chi2}}. 

\begin{table}
\centering
\caption{$\chi^2$ for different models}
\begin{tabular}{c|ccc}
\hline 
\backslashbox{$\epsilon_{\rm small}$}{$s_{\rm big}$} & 0.2 & 0.4 & 0.6 \\ 
\hline 
0.1 & 0.14 & 0.12 & 0.13\\ 
%\hline 
0.01 & 0.16 & 0.19 & 0.23 \\ 
\hline 
\end{tabular} 
\label{Table:chi2}
\end{table}

\begin{figure}
\centering
\includegraphics{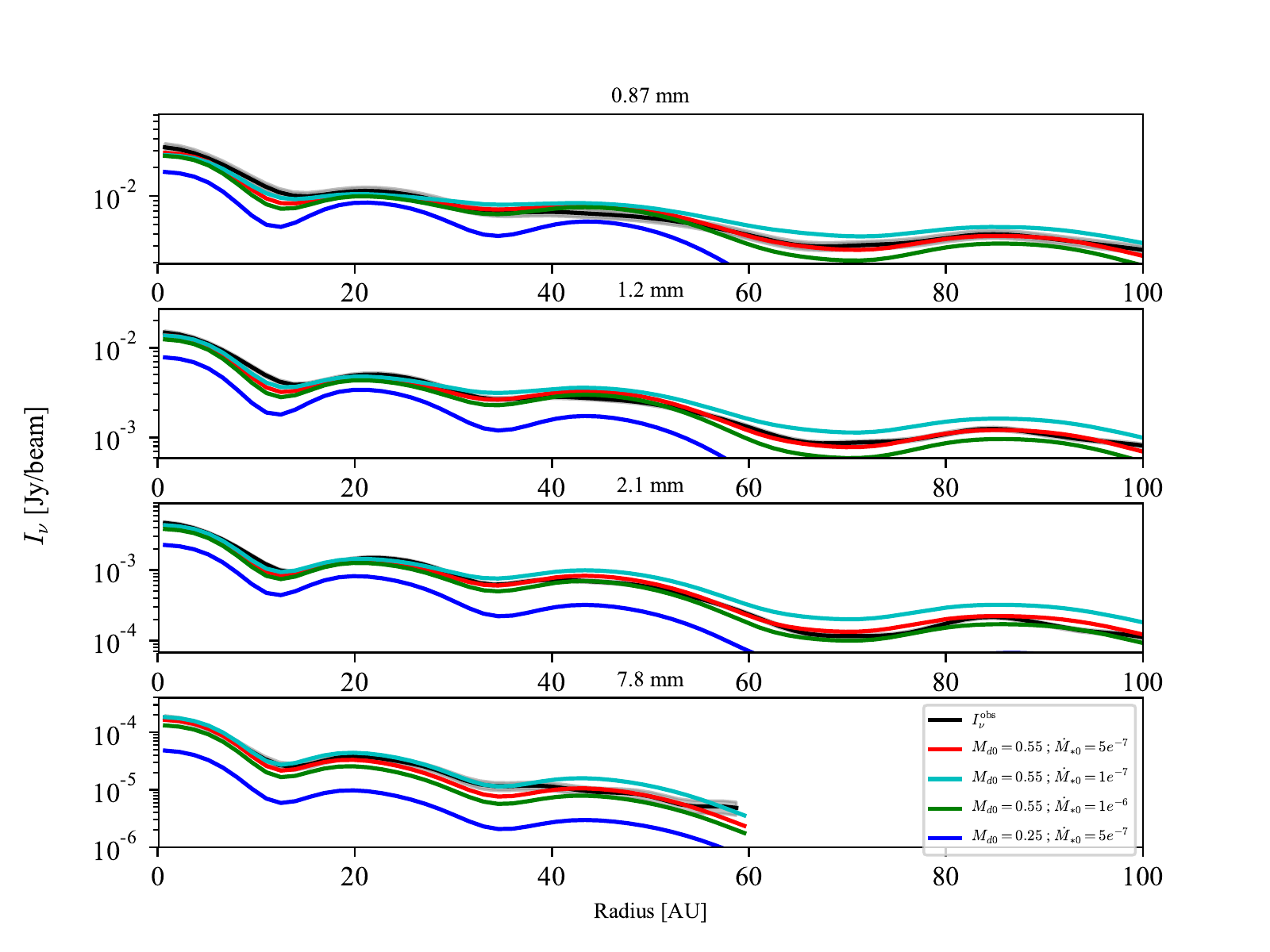}
\caption{
The black solid lines show the observed profiles at at  each wavelength. The different colored lines show models where the
initial disk mass $M_{d0}$ and the initial mass accretion rate $\dot M_{*0}$ vary with respect to 
Model VII, which corresponds to the red solid lines.
 }
\label{Fig:MdMdot}
\end{figure}

Figure \ref{Fig:MdMdot} shows the effect of changing the initial mass accretion rate $\dot M_{*0}$ and disk mass $M_{d0}$ in Model VII. From Equations 
 (\ref{eq:T}) and (\ref{eq:sigma_mod})   one can see that the mass surface density
 increases with $M_{d0}$, and decreases with $\dot M_{*0}$. The viscosity has the opposite dependance 
 on these parameters (see Equation \ref{eq:nu_dim}) and affects the temperature mainly 
 in the active regions located at the disk midplane,  while the stellar irradiation determines the temperature of the
 upper layers (see e.g., Figure 4 of \citealt{Lizano_2016}).
 %(for large $t_{\rm age}$, $\nu$ becomes independent of $M_{d0}$ and $\dot M_{*0}$). 
 In these models the disk emission is mainly affected by changes in the mass surface density, such that 
 the optically thin emission increases with the initial disk mass $M_{d0}$, and,
 for a given $M_{d0}$, the emission decreases with the initial mass accretion rate $\dot M_{*0}$.

Finally, Figure {\ref{Fig:gamma} shows the effect of changing the power-law viscosity exponent to $\gamma=0$, keeping the
same parameters of model VII. In this case the viscosity is independent of radius, 
\begin{equation}
    \nu = \nu_1 =\frac{4}{3} \frac{\dot M_{*0} R_c^2}{M_d(0)} {\mathbb T}^{-1}.
\end{equation}
    The surface density depends on the radius only through the exponential function, 
\begin{equation}
    \Sigma(r, {\mathbb T}) = \frac{M_d(0)}{ \pi R_c^2} {\mathbb T}^{-1/4}  \exp (-r^2),
 \end{equation}
and the mass accretion rate is
 \begin{equation}
  \dot M(r, {\mathbb T})  = \dot M_{*0}  {\mathbb T}^{-5/4}  \exp (-r^2)  \left[ 1 - 4 r^2 \right],
 \end{equation}
where $r= R/R_c$ and $R_c = R_1 {\mathbb T}^{1/2}$.
In addition, the non dimensional time is 
\begin{equation}\label{T}
   {\mathbb T}   = 8   t_{\rm age} \frac{\dot M_{*0}}{M_{d0}} +1 .
\end{equation}
One can see that the model with $\gamma=0$ is flatter than the observed profiles.  

 \begin{figure}
\centering
\includegraphics{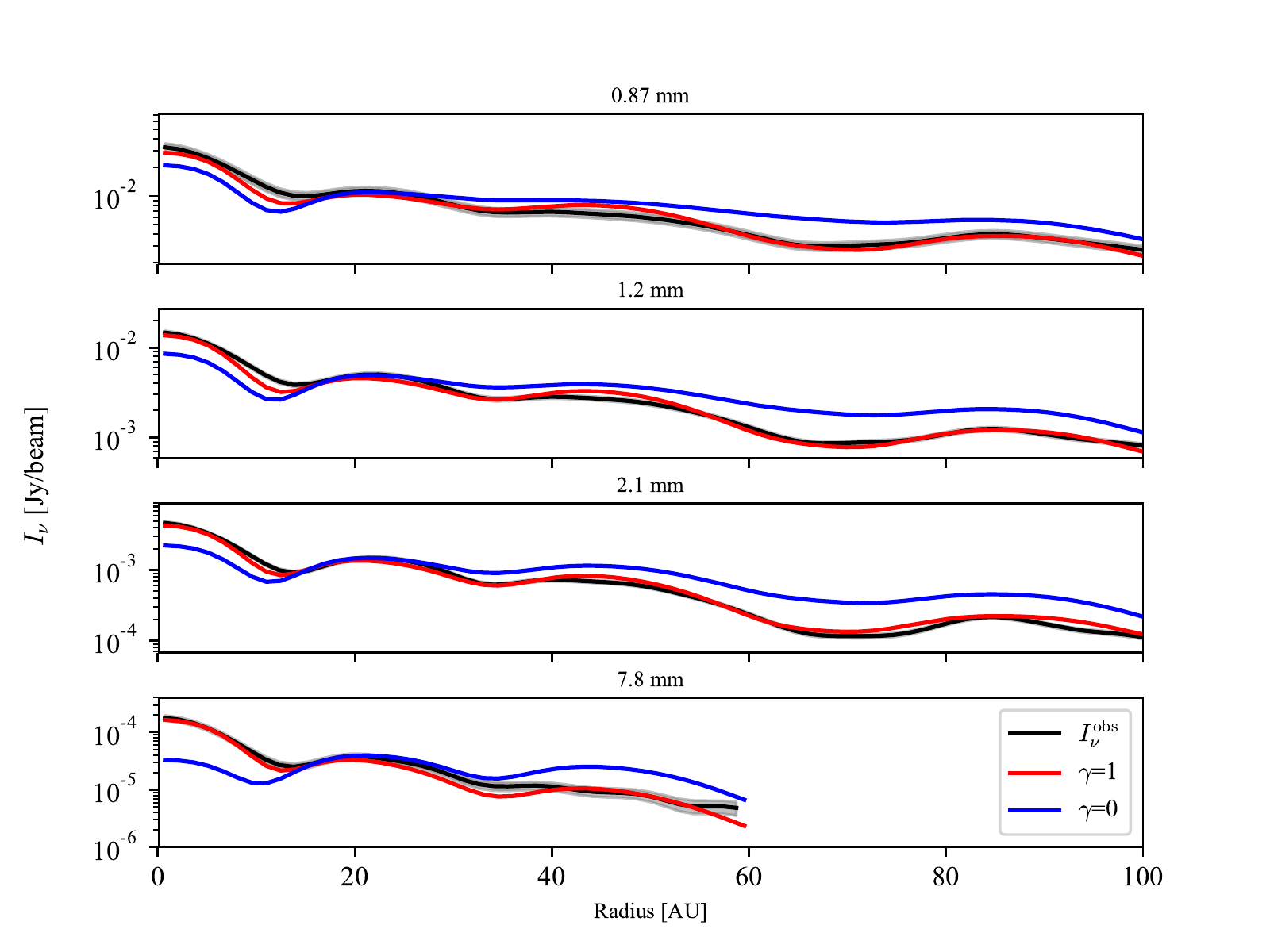}
\caption{
The black solid lines show the observed profiles at at  each wavelength. The red solid lines show the profiles of Model VII.
The blue lines show the profiles of a model with a viscosity power-law exponent $\gamma =0$.
 }
\label{Fig:gamma}
\end{figure}

 \section{Discussion}
\label{SEC:Discussion}
 
 In the previous section we calculate the emission of 
 the Lynden-Bell \& Pringle 
 evolutionary disks with a viscosity power-law exponent
 $\gamma=1$, with different dust properties, in order to reproduce the level of emission of the observed 
 ALMA and VLA profiles. For each model, we adjust the dust ratio $\epsilon$ to obtain the amount of 
 dust mass needed to 
 achieve the observed level of the 7.8 mm VLA emission at the outer disk radii ($40 < R/\au < 60 $). 
 
 We found that Model I, composed of small grains with a size 
 distribution with $a_{\rm max}=100 \ \mu$m in both the atmosphere and the midplane,
  requires a large value of the dust ratio  $\epsilon \sim  9$
 to fit the outer radii of the 7.8 mm profile (Figure \ref{Fig:100mu}). Nevertheless, with this amount of dust, the model profiles at the ALMA wavelengths show excess emission at the inner region of the disk ($R < 20$ au). Thus, this 
  is not a good model to produce the observed dust emission at all 4 wavelengths. 
 
 Models II and III which %a grain size distribution with $a_{\rm max}=100 \ \mu$m and $a_{\rm max}=1$ mm  (see Table \ref{Table:Models})
 have larger grains with with $a_{\rm max}=1$ mm in the midplane, and  $a_{\rm max}=100 \ \mu$m or $a_{\rm max}=1$ mm in the atmosphere
 (see Table \ref{Table:Models}),
also require  a large  dust ratio  $\epsilon=9$ 
to fit the 7.8 mm VLA profile. One can see in Figure \ref{Fig:1mm} that the level of emission in the inner disk 
of the 3 ALMA profiles is obtained only with Model III, where the atmosphere has a grain size distribution with 
$a_{\rm max}=100 \ \mu$m,  since these particles absorb the stellar radiation more efficiently producing a hotter inner region, compared to
Model II . 
%where $a_{\rm max} \sim \lambda/2 \pi$ (e.g., \citealt{Draine_2006}). 
Also,
for the same value of $\epsilon$, 
 both models II and III produce very similar levels of 7.8 mm emission because the small grains in the atmosphere of
Model III do not contribute at this wavelength. Nevertheless, for $R > 50 \ \au$, both models with the $\epsilon=9$,  
have excess emission at ALMA wavelengths: the same dust needed to reproduce the level of the 7.8 mm emission, produces too much emission at these wavelengths.

Models IV, which has $a_{\rm max}=1$ cm in both the atmosphere and the midplane, requires much less dust than the
previous models, $\epsilon=3.5$, to reproduce the level of the 7.8 mm emission for $R > 20$ au. Nevertheless, this model
has a deficit of emission in the inner disk region $ R < 15$ au at all wavelengths (Figure \ref{Fig:1cm}). 

Model V, which has an atmosphere with  $a_{\rm max}=100 \ \mu$m, and the midplane
with $a_{\rm max}=1$ cm, is able to produce the level of emission of all the observed profiles with $\epsilon=3.5$, a
dust-to-gas mass ratio a factor of 3.5 the ISM value. The smaller grains in the atmosphere absorb the stellar radiation
more efficiently than the larger grains of  Model IV, 
heating the inner disk region  and increasing  
the emission at all wavelengths.  In addition, the large cm grains in the disk contribute to the 
7.8 mm emission in the outer disk region, without producing excess emission at the ALMA wavelengths.
Thus, the dust size distribution of Model V is favored over the other the models.

To fit the profiles substructure (bright rings and dark gaps) at ALMA and VLA wavelengths,
we consider variations of Model V where the optical depth is changed in the gaps (Models VI and VII; see Figure \ref{Fig:1cm_gaps}).
Model VI, which has a grain size distribution with $a_{\rm max}=100 \ \mu$m in the atmosphere and in the gaps, and
$a_{\rm max}=1$ cm in the midplane of the rings, has a smaller opacity in the gaps because 
the $100 \ \mu$m particles  have a smaller monochromatic opacity than the 1 cm particles in the rings (see, e.g., Figure 10 of \citealt{Sierra_2017}). 
This model has an excess of emission in the first gap  at ALMA wavelengths because
the gas is hot. It also has excess emission beyond 60 au.
Model VII, which a grain size distribution with $a_{\rm max}=100 \ \mu$m in the atmosphere and
$a_{\rm max}=1$ cm in the midplane, and a deficit of dust mass in the gaps with respect to the rings
($\epsilon(R)$; see Table \ref{Table:ModelVII}), can reproduce the  ring-gap substructure. In fact, a hybrid model with a smaller opacity in the gaps ($a_{\rm max}=100 \ \mu$m) like Model VI and a dust mass deficit in the rings could also reproduce the ring-gap substructure. As discussed above, possibly both effects are at work, given that 
planet formation,  dust trapping in pressure maxima, and sintering due the different volatile snowlines 
will decrease the dust mass and change the grain size in the gaps (e.g., \citealt{Zhang_2018};   \citealt{Sierra_2019};   \citealt{Okuzumi_2019}).

Because of the mass deficit in the gaps, Model VII has a global dust-to-gas mass ratio 
$\zeta_{\rm global}=2 \ \zeta_{\rm ISM}$ within a 100 au. 
Since the gas mass of the disk  is $M_d (100 \ {\rm au}, 1 \ {\rm Myr})= 0.23 M_\odot$ (Equation \ref{eq:mass}),  the dust mass within
this radius is $M^{\rm dust} = \zeta_{\rm global} M_d = 4.8 \times 10^{-3} M_\odot$.
This is the amount of dust required to produce the observed 7.8 mm emission. These values are a factor of 10 times higher than those obtained by \cite{Carrasco-Gonzalez_2016} and CG19, who used a vertically isothermal
structure to infer the disk temperature, in contrast with the vertical structure models considered in this work.
 
We explored the effect of changing the dust settling parameters $s_{\rm big}$ and $\epsilon_{\rm small}$
in Model VII. The differences occur mainly inside $ R \gtrsim 50$ au. One finds that
with a degree of settling  $\epsilon_{\rm small} = 0.1$, the model with the smallest $\chi^2$
 has $s_{\rm big} = 0.4$ (a dust scale height $0.53 H$). 
For a more extreme degree of settling  $\epsilon_{\rm small} = 0.01$,  
the model with the smallest $\chi^2$  has $s_{\rm big} = 0.2$  (a dust scale height $0.25 H$). 
Figure \ref{Fig:Trz} shows radial and vertical temperature structure of these 2 models.  The radial profiles show that
in the inner region, the disk with smaller settling height  ($s_{\rm big} = 0.2$) is colder than the model with
$s_{\rm big} = 0.4$. 
These temperature profiles are comparable to the temperature profile found by \cite{Liu_2017} for the HL Tau disk.
The sharp temperature transitions between rings and gaps are due to the fact that, for simplicity, we 
assumed a sharp change in the dust properties and calculated a 1D vertical structure, which neglects radial gradients. 
For reference, the  dashed black lines show a slope $p=-0.5$. 
The vertical profiles show that, for both models, except for the first ring, the gaps (dashed lines) are warmer than the rings (solid lines). 
The open circles show the location of the settling height $z_{\rm big}$ at each radius, the boundary of the midplane region which has large grains.

\begin{figure}
\centering
\includegraphics[scale=1]{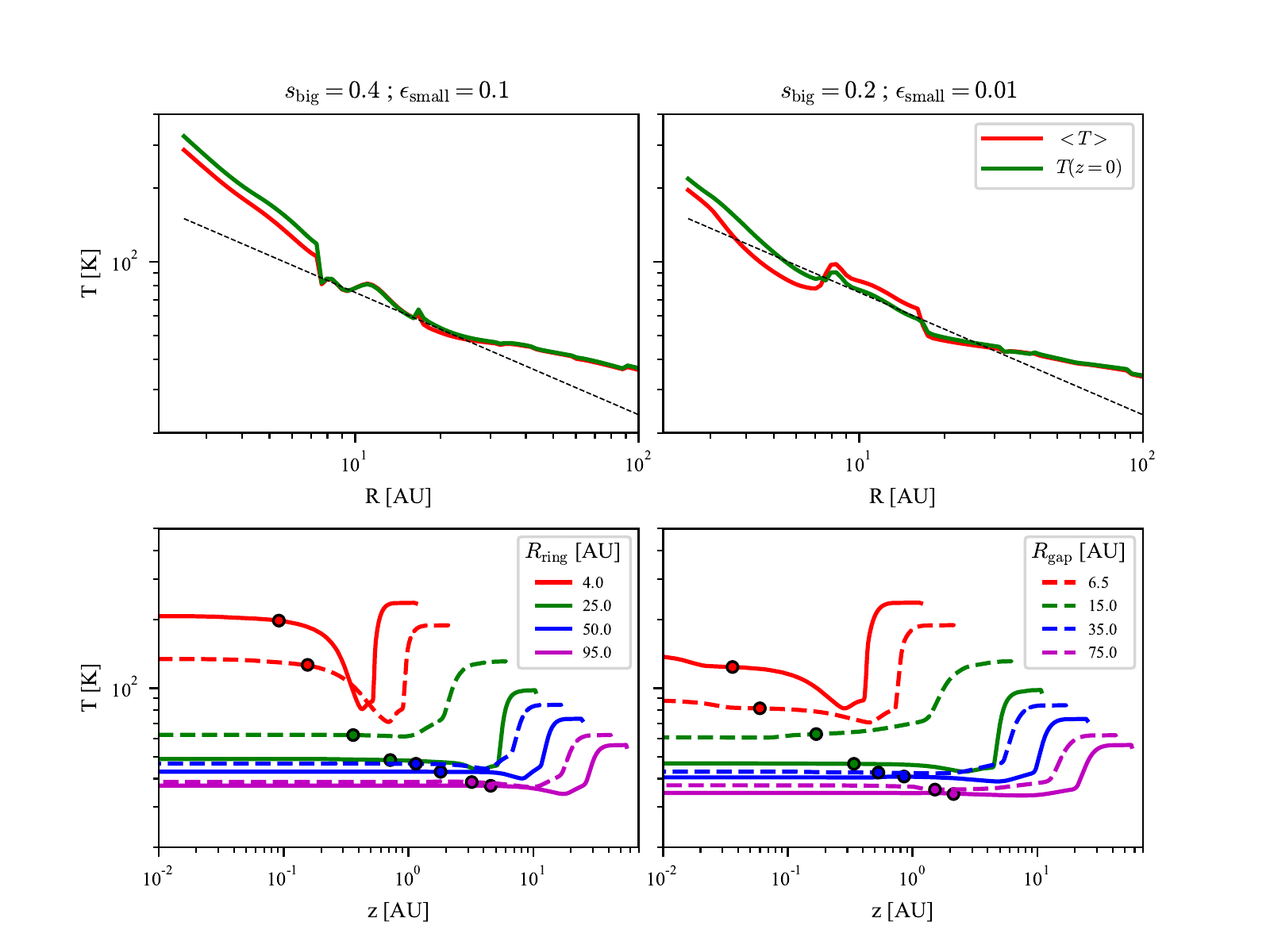}
\caption{Disk structure for $s_{\rm big} = 0.4$ and $\epsilon_{\rm small}$ = 0.1 (left panels); and $s_{\rm big} = 0.2$ and $\epsilon_{\rm small}$ = 0.01 (right panels). The top panels show  the 
average  temperature (red lines) ($<T>= \int T \rho dz / \Sigma_\varpi$) 
and the midplane temperature (green lines) as a function of radius. The black dashed lines have a slope $p=-0.5$. The bottom panels show the temperature as a function of height above the midplane for different radii. The solid lines correspond to radii inside the rings. The dashed lines
correspond to radii inside the gaps. The open dots show the location of settling height $z_{\rm big}$ at each radius.}
\label{Fig:Trz}
\end{figure}

We considered evolutionary disks with a smaller initial disk mass and different
 initial mass accretion rates than Model VII (Figure \ref{Fig:MdMdot}). 
 The emission of these models vary mainly due to the change in the mass surface density
 (see Equations \ref{eq:sigma_mod} and \ref{eq:T}). For example, consider the
 model with a disk initial mass $M_{d0}^\prime =0.25	 \ M_\odot$,
 and an initial mass accretion rate $\dot M_{*0} = 5 \times 10^{-7} M_\odot {\rm yr}^{-1}$,  
which has a disk mass 
$M_d^\prime(100 \ {\rm au}, 1 \ {\rm Myr})= 8.0 \times 10^{-2} M_\odot$,  %(Equation \ref{eq:mass}), 
3 times smaller than Model VII. 
 Given the amount of dust required by Model VII  ($M^{\rm dust} = 4.8  \times 10^{-3} M_\odot$)
 to produce the 7.8 mm emission, one can estimate that this model  would need a global dust-to-gas mass ratio 
$\zeta^\prime = M^{\rm dust} / M_d^\prime = 6 \  \zeta_{\rm ISM} $ to reach the same level of emission.
  
 Finally, we explore a evolutionary disk with the same parameters as Model VII, but with a 
 viscosity power-law exponent $\gamma=0$, closer to the 
 exponent $\gamma=0.2$, used for the functional form of the disk surface density by several authors (e.g., \citealt{Kwon_2015},
\citealt{Pinte_2016}; \citealt{Liu_2017}).  
%\footnote{Nevertheless, note that in models with $\gamma < 1$,  the mass accretion rate decreases too slowly with time compared to the observed decline of
%T Tauri stars of different ages  \citep{Hartmann_1998}.}
Since the initial disk mass is the same in both models, the surface density of the $\gamma=0$ model is lower at the center and higher
at the outer regions, with respect to the $\gamma=1$ model. For this reason, the emission profiles are flatter than the $\gamma=1$ model:
 there is a deficit of emission at the center and an excess of emission at the outer disk regions.
%In addition, the emission at the ALMA wavelengths is low in the outer radii because the viscosity is smaller than the  $\gamma=1$ model. 
%This model can reproduce the 7.8 mm VLA emission  at radii $R > 30$ au 
%with the same dust mass as Model VII, i.e., $M^{\rm dust} = 5.05 \times 10^{-3} M_\odot$, 
%although the model parameters could be adjusted to fit better the observed profiles (e.g., a different function $\epsilon(R)$).
%In fact, it would be important to obtain high spatial resolution and high sensitivity maps 
%of the gas emission of HL Tau  (e.g., CO isotopes)  to obtain a physically motivated
%dust ratio function $\epsilon(R)$, for example, with a model of dust radial migration (e.g., \citealt{Sierra_2019}).

\section{Conclusions}
\label{SEC:Conclusions}
We probe the dust grain properties in the disk of HL Tau using the disk evolutionary models of Lynden-Bell \& Pringle
 with a viscosity power-law exponent $\gamma=1$ by
 comparing the models' emission with multi-wavelength ALMA and VLA profiles. 
This comparison favors disks with a dust
size distribution with small grains in the atmosphere ($a_{\rm max} = 100 \ \mu$m) and large grains in the midplane
($a_{\rm max} = 1$ cm). 

The bright and dark ring substructure in the  observed profiles can be reproduced by changing the optical depth in
the gaps. This can be done by either a change in opacity (small grains in the gaps with $a_{\rm max} = 100 \ \mu$m) 
or a dust mass deficit in the gaps (a dust ratio $\epsilon(R)$ larger in the rings than in the gaps), or a combination of both.
The model that reproduces the multi-wavelength ALMA and VLA profiles of HL Tau
 has a global dust-to-gas mass ratio 2  the ISM value and 
 contains a total dust mass $M^{\rm dust} = 4.8 \times 10^{-3} M_\odot$   within 100 au. 
This dust mass is required to achieve the emission level of the 7.8 mm VLA profile.

%We explore the effect on the disk emission of changing the initial disk mass and the mass accretion rate and find that
%the dust ratio required to reproduce the observed level of emission at 7.8 mm increases 
%for values of the parameters that decrease the mass surface density. Also, varying the surface density of the midplane region $s_{\rm big}$
%and the degree of dust settling $\epsilon_{\rm small}$ can vary the emission by factors of less than 2.
In the evolutionary models, decreasing the initial disk mass  or increasing the initial mass accretion rate with respect to the fiducial model, 
decreases the mass surface density and thus, decreases the emission at 7.8 mm. 
A dust scale height between  0.3 - 0.5  of the gas scale height $H$ can fit the observed profiles.
%Also, varying the surface density of the midplane region
%with large grains  %$s_{\rm big}$ 
%and/or  the degree of dust settling, %$\epsilon_{\rm small}$, 
%changes the emission by less than a factor of 2.
In addition, the millimeter emission profiles of a model with a mass surface density with a power-law exponent 
$ \gamma =0$,  are flatter  than the observed ALMA and VLA profiles.

Although we focused on a specific evolutionary model  where the disk structure and emission can be
calculated self-consistently,  we conclude from this work
 that the high resolution optically thin millimeter (VLA)  observations are very important to determine the dust properties and dust mass in
protoplanetary disks. With such data, it will be possible to infer the dust properties of many more 
protoplanetary disks in the near future.

\bigskip
\textit{Acknowledgements.}
C. T., S. L., A. S., and E. B-B 
acknowledge support from PAPIIT-UNAM IN101418 and CONACyT 23863.

\appendix
\section{Dust settling}
\label{App:1}

In this appendix we rewrite the formalism of  D'Alessio et al. (2006) in terms of the gas surface density, including a dust-to-gas mass ratio that can be a function of radius $\zeta(R)$.
\footnote{ $\zeta(R)$ can be the ISM value $\zeta_{\rm ISM} = 1/100$ or another value that takes into account, e.g., dust radial migration and/or dust depletion in gaps.}
These authors assume that the dust settles to the midplane and grows. 
Then, the disk has a distribution of small grains in the upper layers (e.g, $a_{\rm max} = 1- 100 \ \mu$m) and a distribution of big grains 
(e.g, $a_{\rm max} =$ 1 mm - 1 cm) in the midplane.  At each radius, the total gas mass surface density is 
$\Sigma(R) = 2 \int_0^\infty \rho dz$ where $\rho$ is the volume density. and $z$ is the height over the disk midplane.

The total surface density can be written in terms of the surface density of the midplane layers with big grains $\Sigma_{\rm big}= 2\int_0^{z_{\rm big}} \rho d z $, where $z_{\rm big}$ is the height of the midplane region, 
and the surface layers with small grains $\Sigma_{\rm small} = 2\int_{z_{\rm big} }^\infty \rho d z$,
\begin{equation}
\Sigma(R) = \Sigma_{\rm big} + \Sigma_{\rm small} .
%\equiv 2 \left( \int_0^{z_{\rm big}} \rho d z + \int_{z_{\rm big} }^\infty \rho d z \right) .
\label{Sigma_varpi}
\end{equation}

Originally, the surface and midplane layers had the same dust-to-gas mass ratio $\zeta(R)$. 
Due to settling, the surface layers lost dust such that they now have
a smaller dust-to-gas mass ratio $\zeta_{\rm small}$. Then, the dust mass lost by the upper layers is given by 
the original dust mass minus the dust mass the remains in these layers
\begin{equation}
\Sigma_{\rm up, lost}^{d} = \left( \zeta(R) - \zeta_{\rm small} \right) \Sigma_{\rm small} .
\end{equation}
The midplane layers have a dust surface density given by the original dust mass plus the mass of settled grains from the upper layers,
\begin{equation}
\Sigma_{\rm down}^{d} = \left[ \zeta(R) \Sigma_{\rm big} + \left (\zeta(R) - \zeta_{\rm small} \right) \Sigma_{\rm small} \right]
\equiv  \zeta_{\rm big} \Sigma_{\rm big} .
\label{zeta_big}
\end{equation}
This equation defines the midplane dust-to-gas mass ratio $\zeta_{\rm big}$.
This equation can be rewritten in terms of the dust ratios $\epsilon(R)= \zeta(R) /\zeta_{\rm ISM}$, $\epsilon_{\rm small} = \zeta_{\rm small} / \zeta_{\rm ISM}$,
and $\epsilon_{\rm big} = \zeta_{\rm big} / \zeta_{\rm ISM}$ as
\begin{equation}
\epsilon_{\rm big} = \epsilon(R)  + \left( \epsilon(R) - {\epsilon_{\rm small} } \right)  
\frac{\Sigma_{\rm small}}{\Sigma_{\rm big}} 
= \epsilon(R)  + \left( \epsilon(R) - {\epsilon_{\rm small} } \right)  
\left( \frac{\Sigma(R) }{\Sigma_{\rm big}} - 1\right),
\end{equation}
where the last equality comes from Equation (\ref{Sigma_varpi}). In the case $\zeta(R) = \zeta_{ISM}$,
$\epsilon (R) =1$, and
\begin{equation}
\epsilon_{\rm big} =  1 + \left( 1 - {\epsilon_{\rm small} } \right)  
\left( \frac{\Sigma(R) }{\Sigma_{\rm big}} - 1\right).
\label{epsilon_big_ISM}
\end{equation}
For example, for $\epsilon_{\rm small} = 0.1$, and 
$\Sigma_{\rm big} / \Sigma(R) = 0.6$, this equation gives $\epsilon_{\rm big} = 1.6$. 

To avoid problems of convergence in the code of the vertical structure,
one makes a smooth transition between the upper and lower layers and calculates the dust ratio $\epsilon$ as a function of the surface density measured from the midplane 
 $\sigma = 2 \int_0^z \rho dz$ as

\begin{equation}
\epsilon_{\rm small} (\sigma)= \frac{\zeta_{\rm small} (\sigma) }{\zeta_{\rm ISM}} = 0.5 \, \epsilon_{ \rm small} \, \prn{1-\tanh \brk{k \prn{1- \frac{\sigma }{\sigma_{\rm big}}}}},
\end{equation}

and
\begin{equation}
\epsilon_{\rm big} (\sigma)= \frac{\zeta_{ \rm big} (\sigma)}{\zeta_{\rm ISM}} = 0.5 \, \epsilon_{ \rm big} \, \prn{1+\tanh \brk{k \prn{1- \frac{\sigma}{\Sigma_{\rm big}}}}},
\end{equation}
where %$\Sigma_{\rm big} / \Sigma(R) < 1$, and  
$0 < \sigma < \Sigma(R) $. 

%Figure 1 shows the functions $\epsilon_{\rm big} (\Sigma)/ \epsilon_{\rm big} $ and  $\epsilon_{\rm small} (\Sigma)/ \epsilon_{\rm small} $. 
%Figure 2 shows $\zeta_{\rm big}(\Sigma)/\zeta_{\rm ISM}$ and $\zeta_{\rm small}(\Sigma)/\zeta_{\rm ISM}$.

%\begin{figure}
%\centering
%\includegraphics{Transitions.pdf}
%\caption{Caption}
%\label{FIG:Transtition}
%\end{figure}

%\begin{figure}
%\centering
%\includegraphics{Transitions_2.pdf}
%\caption{Caption}
%\label{FIG:Transtition}
%\end{figure}

%To obtain an approximate relation between $\Sigma_{\rm big}$ and $z_{\rm big}$}

In addition, the relation between the  settling height $z_{\rm big}$ and the midplane mass surface density $\Sigma_{\rm big}$ can be easily obtained 
for a vertically isothermal disk where the density is given by
\begin{equation}
\rho = \rho_0 \exp{\left ( \frac{-z^2}{2 H^2}  \right)} ,
\end{equation}
where $z$ is the vertical coordinate and $H$ is the disk scale height. In this case, the total gas mass surface density
is given by
\begin{equation}
\Sigma(R) = 2 \int_0^\infty \rho_0 \exp{\left ( \frac{-z^2}{2 H^2}  \right)} dz 
% = 2 \rho_0 \sqrt{2} H \int_0^\infty \exp{(-x^2)} dx
= \sqrt{2 \pi} H \rho_0 .
\end{equation}
If one writes $z_{\rm big} = \delta H$, 
\begin{equation}
\Sigma_{\rm big} = 2\int_0^{\delta H} \rho_0 \exp{\left ( \frac{-z^2}{2 H^2}  \right)} dz 
= \sqrt{2\pi} H \rho_0  \erf{\left( \frac{\delta}{\sqrt{2}}  \right)} .
\end{equation}
Then, the factor $\delta$ is given by the trascendental equation
\begin{equation}
\frac{\Sigma_{\rm big}}{\Sigma(R) } = \erf{\left( \frac{\delta}{\sqrt{2}}\right) }.
\end{equation}
For example, for ${\Sigma_{\rm big}}/{\Sigma(R) }  = 0.6 $ and 0.2,   %$\delta = 0.25$ and 
$z_{\rm big} = 0.84 H$ and $0.25 H$, respectively.  

For non isothermal disks, the height coefficient $\delta$ can be found directly from the disk vertical structure, evaluating the 
scale height at the midplane temperature.

%If ${\Sigma_{\rm big}}/{\Sigma(R) }  = 0.9$ then $\delta = 1.67$.

% \input{References.tex}

\end{document}